\documentclass[twocolumn,showpacs,preprintnumbers,amsmath,amssymb]{revtex4}
\usepackage{bm,amsmath,amssymb,latexsym,mathrsfs,graphicx,enumerate}
\usepackage[mathcal]{euscript}
\usepackage{hyperref}
\usepackage{epsfig}

\DeclareMathOperator{\sign}{sgn}

\begin{document}

\title{\textbf{Nonlinear waves in a positive-negative coupled waveguide zigzag array}}
\author{Elena V. Kazantseva$^{1,3}$,  Andrey I. Maimistov$^{2,3}$}
\affiliation{\normalsize \noindent $^1$: Joint Institute of High
Temperatures, RAS, Moscow,
125412, Russia \\
$^2$: Department of General Physics, Moscow Institute for Physics
and Technology, Dolgoprudny, Moscow region, 141700 Russia \\
$^3$: Department of Solid State Physics and Nanostructures, National
Nuclear Research University
Moscow Engineering Physics Institute, Moscow, 115409  \\
E-mails: elena.kazantseva@gmail.com, ~~
aimaimistov@gmail.com \\
}
\date{\today}

\begin{abstract}
\noindent We consider the coupled electromagnetic waves propagating
in a waveguide array, which consists of alternating waveguides of
positive and negative refraction indexes.  Due to zigzag
configuration there are interactions between both nearest and next
nearest neighboring waveguides exist. It is shown that there is a
stop band in the spectrum of linear waves. The system of evolution
equations for coupled waves has the steady state solution describing
the electromagnetic pulse running in the array. Numerical simulation
demonstrates robustness of these solitary waves.

\end{abstract}



\maketitle

\section{Introduction}

\noindent Recently new applied physics field is referred to as
transformation optics develops
\cite{Bergamin:08,Urzhumo:Smith:10,KGSmith:10,Shal:Pendry:11,Kild:Shal:11,Diedrich:12}.
The transformation optics owes its origin to the artificial
materials --- metamaterials, which are featured by unusual
electrodynamics properties
\cite{Ch:Wu:Kong:06,Boltasseva:08,Elefth:05,Noginov:12}. For
instance these media are characterized by a negative refraction
index (NRI) when the real parts of the dielectric permittivity and
the magnetic permeability are simultaneously negative in a certain
frequency range.

The negative refraction in metamaterials manifest themselves when
the wave passes through the interface between such medium and a
conventional dielectric or in the case of the surface waves
\cite{Darmanyan:05,Shadrirov:04}. The novel kind of nonlinear
interaction of the guided waves can be realized in a nonlinear
oppositely-directional coupler \cite{MGL:08,KMO:09}. This coupler
consist of two tightly spaced nonlinear/linear waveguides. The sign
of the index of refraction of one of these waveguides is positive
and the index of refraction of other waveguide is negative. The
opposite directionality of the phase velocity and the energy flow in
the NRI waveguide facilitates an effective feedback mechanism that
leads to optical bistability ~\cite{LGM:07} and gap soliton
formation \cite{KMO:09,Rizhov:Maim:12}.

\begin{figure}
\includegraphics[width=0.4\textwidth]{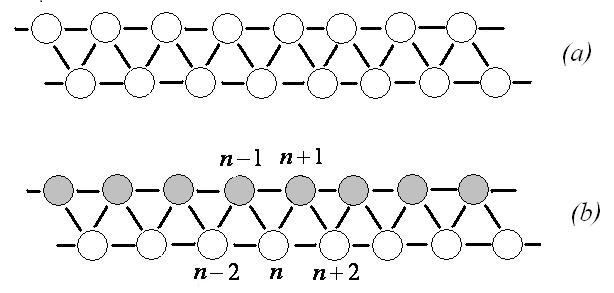}
\caption{{\protect\footnotesize {\ A schematic illustration of
zigzag coupled waveguides array (a); and zigzag positive-negative
coupled waveguides array (b).}}} \label{zigzag:1}
\end{figure}

The optical waveguide array provide a convenient setup for
experimental investigation of periodic nonlinear systems in one
dimension \cite{Christod:Leder:Silber:03}. Nonlinear optical
waveguide arrays (NOWA) are a natural generalization of nonlinear
couplers. NOWA with a positive refractive index have many useful
applications and are well studied in the literature (see for
example~\cite{Christo:88,Sch:Lederer:91,Darmanyay:Leder:97}). If the
sign of the index of refraction  of one of waveguides in NOWA is
positive and the index of refraction of other neighboring waveguide
waveguide is negative the alternated NOWA will be obtained
\cite{Maim:Gabi:2007,Konotop:Abdullaev:12,Maim:Kazant:Desyat:12}.

Usually the coupling between nearest neighboring waveguides is
taken into account. It is correct approximation for strong
localized electromagnetic wave in waveguide. However, the coupling
between both nearest neighboring waveguides and the next nearest
neighboring ones can be introduced by the use of a zigzag arrangement
~\cite{Efre:Chri:02,Wang:Huang:Yu:10}. Let $\vartheta_b$ is an
 angle between the lines connecting the centers of neighboring
waveguides. In a linear array this angle is $\pi$. In a zigzag like
array at $\vartheta_b\approx \pi/2$ the coupling between the nearest
neighboring waveguides and the next nearest neighboring ones is
approximately the same. Nonlinear optical waveguide zigzag arrays
(NOWZA) can be considered as generation of NOWA.

In this paper we will consider nonlinear optical waveguide array that
looks like zigzag array discussed in \cite{Efre:Chri:02}. However in
the case under consideration the refractive indexes of the
neighboring waveguides have a different sign (Fig. 1 b). The
dispersion relation of the linear wave will be determined. We
suppose that only waveguides with PRI are manufactured from cubic
nonlinear dielectric. Steady state solitary waves of the particular
kind will be found. Numerical simulation of the steady state waves
collision will be used to illustrate the robustness of these solitary
waves.

\section{The model formulation}

\noindent Let us assume that waveguide marked by index $n$ is
characterized by positive refractive index (PRI), the nearest
neighboring waveguides with indexes $n-1$ and $n+1$ possess negative
refractive index. If the electromagnetic radiation is localized
in each waveguide the coupled wave theory can be used. In the case
of the array which is prepared in the form of line (i.e., all centers of
waveguide placed on one line), only the interaction between the
nearest neighboring waveguides is essential. In this configuration
the angles between the lines connecting waveguides are equal $\pi$.
However, if the array is deformed in the form of zigzag, where the
angles between the lines connecting waveguides are equal
approximately $2\pi/3$, interaction between both the nearest
neighboring and the next nearest neighboring waveguides will be
important \cite{Efre:Chri:02}.

The system of equations describing the wave propagation in a
nonlinear zigzag waveguide array with alternated sign of refractive
index reads as
\begin{eqnarray}
&& i\left(\frac{\partial}{\partial \zeta} +\frac{\partial}{\partial
\tau} \right)e_n + c_1(e_{n+1}+e_{n-1})+\notag \\
&&\qquad + c_2(e_{n+2}+e_{n-2}) +r_1|e_n|^2e_n =0, \label{km:1:1} \\
&& i\left(-\frac{\partial}{\partial \zeta} +\frac{\partial}{\partial
\tau} \right)e_{n+1} + c_1(e_{n+2}+e_{n})+\notag \\
&& \qquad +c_3(e_{n+3}+e_{n-1}) r_2|e_n|^2e_n =0,  \label{km:1:2}
\end{eqnarray}
where $e_n$ is the normalized envelope of the wave localized in
$n$-th waveguide. Coupling between neighboring PRI and NRI
waveguides is defined by parameter $c_1$. The parameter $c_2$
($c_3$) is coupling constant between neighboring PRI (NRI)
waveguides. The all functions $e_{n}(\zeta,\tau)$, independent
variables $~\zeta,~\tau$ and other parameters are expressed in
terms of the physical values represented in
\cite{Maim:Gabi:2007}. We will assume that NRI waveguides are linear
ones ($r_2=0$). The configuration of these alternating waveguides
will be remarked as asymmetric alternating nonlinear optical
waveguide zigzag arrays (ANOWZA).

\section{Linear properties of the alternating waveguide zigzag array}

\noindent let us consider the case of the linear waveguides combined
into zigzag alternating waveguide array. This optical system is
described by following equations
\begin{eqnarray}
&& i\left(\frac{\partial}{\partial \zeta} +\frac{\partial}{\partial
\tau} \right)e_n + c_1(e_{n+1}+ e_{n-1})+ \notag \\
&& \qquad\qquad \quad +c_2(e_{n+2}+e_{n-2})  =0, \label{km:37:1}\\
&& i\left(\frac{\partial}{\partial \zeta} -\frac{\partial}{\partial
\tau} \right)e_{n+1} - c_1(e_{n+2}+e_{n}) -\notag \\
&& \qquad\qquad \quad - c_3(e_{n+3}+e_{n-1}) =0. \label{km:37:2}
\end{eqnarray}

To find the linear wave spectrum we can employ the presentation of
the envelopes in the form of harmonic waves
\[e_n=Ae^{-i\omega\tau +iq\zeta +i\phi n}, \quad
e_{n+1}=Be^{-i\omega\tau +iq\zeta +i\phi(n+1)}.
\]
Substitution of this expression in (\ref{km:37:1}) and
(\ref{km:37:2}) leads to a system of the algebraic linear equations
respecting $A$ and $B$.  Nonzero solutions of these equations exist if the determinant
\[\det \left(
\begin{array}{cc}
  q+\omega + \gamma_3 & \gamma_1 \\
  -\gamma_1 & q -\omega - \gamma_2 \\
\end{array}\right)
\]
is equal to zero. Here the parameters $\gamma_1 = 2c_1\cos \phi$,
$\gamma_2 = 2c_2\cos 2\phi$, and $ \gamma_3 = 2c_3\cos 2\phi$ were
introduced. This condition results in equation
\[(\omega+\omega_0)^2 = \gamma_1^2+(q-q_0)^2\]
where \[2\omega_0 = (\gamma_2+\gamma_3), \quad 2q_0 =
(\gamma_2-\gamma_3).
\]
Thus, we obtain the dispersion law for linear waves in linearized
ANOWZA
\begin{equation}\label{k:m:40}
    \omega^{(\pm)}(q) = -\omega_0 \pm \sqrt{\gamma_1^2 + (q-q_0)^2}.
\end{equation}
Expression (\ref{k:m:40}) shows that  (a) the forbidden zone (gap)
in spectrum of the linear waves exist $\Delta\omega = 2\gamma_1$,
(b) spectrum is shifted along both frequency axis and wave numbers
axis, (c) the form of spectrum likes the spectrum for linear
oppositely-directional coupler
\cite{KMO:09,Maim:Gabi:2007,Maim:Kazant:Desyat:12}. The gapless
spectrum appears only when condition $\phi=\pi/2$ is hold. In this
case the radiation propagates along waveguides with the same
refractive indexes. Energy flux between neighboring waveguides is
zero.

\section{Nonlinear waves in ANOWZA}

\noindent The simplest approximation is based on following ansatz
\[e_{n}(\zeta,\tau)=A(\zeta,\tau)e^{i\phi n}, \quad
e_{n+1}(\zeta,\tau)=B(\zeta,\tau)e^{i\phi(n+1)},
\]
where $A$ and $B$ are the envelopes of the quasi-harmonic waves. It
allows to do reduction of the equations (\ref{km:1:1})-(\ref{km:1:2})
and to obtain the equations
\begin{eqnarray}
&& i\left(\frac{\partial}{\partial \zeta} +\frac{\partial}{\partial
\tau} \right)A + \gamma_1B +\gamma_2A +r|A|^2A =0, \label{km:42:1}\\
&& i\left(\frac{\partial}{\partial \zeta} -\frac{\partial}{\partial
\tau} \right)B - \gamma_1A- \gamma_3B =0. \label{km:42:2}
\end{eqnarray}

At $\phi=\pi/2$ this system of equations is splitting in two
independent equations. At $\phi=\pi/4$ (\ref{km:42:1})-(\ref{km:42:2})
is transformed to that, which was considered in
\cite{MGL:08,KMO:09}.

Real amplitudes and phases can be introduced by means of the
formulae $A=a_1\exp(i\varphi_1)$ and $B=a_2\exp(i\varphi_2)$. The
system of equations (\ref{km:42:1}) and (\ref{km:42:2}) can be
represented in the real form:
\begin{eqnarray*}
  && \left(\frac{\partial}{\partial \zeta} +\frac{\partial}{\partial
\tau} \right)a_1 = \gamma_1a_2\sin\Phi, \\
&&\left(\frac{\partial}{\partial \zeta} -\frac{\partial}{\partial
\tau} \right)a_2 = \gamma_1a_1\sin\Phi, \\
  &&\left(\frac{\partial}{\partial \zeta} +\frac{\partial}{\partial
\tau} \right)\varphi_1 = \gamma_1\frac{a_2}{a_1}\cos\Phi +\gamma_2
+r a_1^2,\\
  && \left(\frac{\partial}{\partial \zeta} -\frac{\partial}{\partial
\tau} \right)\varphi_2 = -\gamma_1\frac{a_1}{a_2}\cos\Phi -\gamma_3,
\end{eqnarray*}
where $\Phi = \varphi_1-\varphi_2$.

\subsection{Steady state waves}

\noindent Among the different kinds of the waves the steady state
waves attract attention. These waves correspond to solution of the wave
equation depending only one particular variable. Let this variable
be
\[\xi = \gamma_1 \frac{\zeta +\beta \tau}{\sqrt{1-\beta^2}},\]
where $\beta$ is parameter that defines the group velocity. If we
introduce new envelopes $u_1$ and $u_2$, according to formulae
$u_1=\sqrt{1+\beta}a_1$ $u_2=\sqrt{1-\beta}a_2$, then $u_1$, $u_2$
and $\Phi$ are governed by following equations
\begin{eqnarray}
  \frac{\partial u_1}{\partial \xi}&=& u_2 \sin\Phi, \label{k:m:28:1} \\
  \frac{\partial u_2}{\partial \xi}&=& u_1 \sin\Phi, \label{k:m:28:2} \\
  \frac{\partial \Phi}{\partial \xi} &=& \delta+\left(\frac{u_1}{u_2}+ \frac{u_2}{u_1}
  \right)\cos\Phi + \vartheta u_1^2, \label{k:m:28:3}
\end{eqnarray}
where
\begin{eqnarray*}
  && \vartheta =
\frac{r}{\gamma_1(1+\beta)}\sqrt{\frac{1-\beta}{1+\beta}}, \\
  && \delta = \left(\frac{\gamma_3}{\gamma_1}
\sqrt{\frac{1+\beta}{1-\beta}}+\frac{\gamma_2}{\gamma_1}
\sqrt{\frac{1-\beta}{1+\beta}}\right).
\end{eqnarray*}

From equations (\ref{k:m:28:1}) and (\ref{k:m:28:2}) the first
integral of motion follows
\begin{equation}\label{k:m:28:integral:1}
    u_1^2-u_2^2 = C_1 =\mathrm{const}.
\end{equation}
If both parts of (\ref{k:m:28:3}) multiply on $u_1u_2\cos\Phi$ and
take into account (\ref{k:m:28:1}) and (\ref{k:m:28:2}), then the
second integral of motion can be obtained
\begin{equation}\label{k:m:28:integral:2}
    u_1u_2\cos\Phi +\frac{\delta}{2}u_1^2 +\frac{\vartheta}{4}u_1^4 = C_2 =\mathrm{const}.
\end{equation}

In depending on the boundary conditions the different solutions of
the equations (\ref{k:m:28:1})--(\ref{k:m:28:3}) can be found. Any
of these solutions will be describing the steady state waves.

\subsection{Solitary waves}

\noindent Solutions of the equations
(\ref{k:m:28:1})--(\ref{k:m:28:3}) under boundary conditions
$u_{1,2} \to 0$ at $|\xi| \to \infty$ correspond with steady state
solitary waves. According to the boundary conditions the constants
 $C_1$ and $C_2$ of (\ref{k:m:28:integral:1})
and (\ref{k:m:28:integral:2}) are zero. Hence
\[u_1^2 -u_2^2=0, \quad \varepsilon\cos\Phi +\frac{\delta}{2}
+\frac{\vartheta}{4}u_1^2 = 0, \quad \varepsilon =\pm 1.\] The
system of equations (\ref{k:m:28:1})--(\ref{k:m:28:3}) with taking
into account  $u_{2}= \varepsilon u_{1}$ can be reduced to the following
one
\begin{eqnarray*}
  \frac{\partial u_1}{\partial \xi}&=& \varepsilon u_1 \sin\Phi, \\
  \frac{\partial \Phi}{\partial \xi} &=& \delta +2\varepsilon \cos\Phi+ \vartheta
  u_1^2.
\end{eqnarray*}
Taking into account the second integral of motion the equation for
$u_1$ is represented as
\[\left(\frac{\partial u_1}{\partial \xi}\right)^2
=u_1^2\left[1-\left(\frac{\delta}{2} +\frac{\vartheta}{4}u_1^2
\right)^2\right].\] Substitution $u_1=w^{-1/2}$ and some little
manipulations leads this equation to
\[\left(\frac{\partial w}{\partial
\xi}\right)^2 =4\Delta^2(w-w_1)(w+w_2), \] where
\[\Delta^2=\left(1-\frac{\delta^2}{4}\right), ~~ w_1 =
\frac{\vartheta}{4(1-\delta/2)}, ~~ w_2 =
\frac{\vartheta}{4(1+\delta/2)}.
\]
Integration of the last equation for $w$ can be done (see
\cite{KMO:09}). It results in \[w = w_0 + w_3\cosh[2\Delta(\xi -
\xi_0)],\] where $\xi_0$ is a constant of integrating. Thus, we have
\[w_0 =\frac{\delta\vartheta}{4(1-\delta^2/4)}, \quad
w_3 =\frac{|\vartheta|}{4(1-\delta^2/4)}.
\]

In term of original variables the functions $u_{1,2}^2(\xi)$ can be
written as
\begin{equation}
u_1^2(\xi) = u_2^2(\xi) =
\frac{4\Delta^2/|\vartheta|}{\cosh[2\Delta(\xi - \xi_0)]+\delta/2}.
\label{k:m:solito:gen}
\end{equation}

So, the real envelopes of the solitary wave propagating in ANOWZA
are presented by following expressions
\begin{eqnarray}
  a_1^2(\xi) &=& \frac{4\Delta^2}{|\vartheta|(1+\beta)\{\cosh[2\Delta(\xi -
\xi_0)]+\delta/2\}}, \label{k:m:solito:1}\\
  a_2^2(\xi) &=& \frac{4\Delta^2}{|\vartheta|(1-\beta)\{\cosh[2\Delta(\xi -
\xi_0)]+\delta/2\}}, \label{k:m:solito:2}
\end{eqnarray}

The phase difference $\Phi$ evolves according to expression
\[\Phi(\xi) = \Phi(-\infty) + \sign(\vartheta)\mathcal{S}(\delta/2;X_0),\]
where auxiliary function
\[\mathcal{S}(a;X_0)=\arctan \frac{(1-a^2)^{1/2} e^{X_0}}{1+a
e^{X_0}},\] and the argument $X_0=2\Delta(\xi - \xi_0)$ were
introduced.

Due to that at $\xi\to -\infty$ the derivative $\partial u_1/\partial \xi$ to
be positive, the value of the $\Phi(-\infty)$ has to equal
$\varepsilon\pi/2$, where $\varepsilon = \pm 1$.

\subsection{Algebraic solitary waves}

\noindent The solutions of the equations
(\ref{k:m:28:1})-(\ref{k:m:28:3}) describe the exponentially
decaying wave fronts. However some times the solitary waves can be
decreasing as $\sim 1/\xi^2$.

The solutions (\ref{k:m:solito:1})-(\ref{k:m:solito:2}) are
characterized by parameter $\Delta$, which can be set to zero in limit
of $|\delta| \rightarrow 2 $. Since, the solutions found here are correct
if $-2 <\delta <2$. However, on the boundaries of this interval we
have to refine behavior of the solitary waves. Let us consider the
function $u^2(\xi)$ near little values of $\Delta$. At $\Delta \ll
1$ from (\ref{k:m:solito:gen}) it follows
\begin{eqnarray*}
  && u^2(\xi) \approx \frac{4\Delta^2/\vartheta}{1+2\Delta^2(\xi -
    \xi_0)^2 +\delta/2} = \\
  && = \frac{4}{\vartheta}\left( \frac{(1-\delta^2/4)}{1+\delta/2 + 2 (1-\delta^2/4)(\xi -
    \xi_0)^2}\right) = \\
  && =  \frac{4}{\vartheta} \left(\frac{(1-\delta/2)(1+\delta/2)}
    {(1+\delta/2) + 2 (1-\delta/2 )(1+\delta/2)(\xi -
    \xi_0)^2}\right) = \\
  && = \frac{4}{\vartheta} \left(\frac{1-\delta/2}{1+2(1-\delta/2)(\xi -
    \xi_0)^2}\right)
\end{eqnarray*}

Near left boundary of the interval for permissible value of
parameter $\delta$, the function  $u^2(\xi)$ can be found using the
substitution $\delta= -2+\varepsilon$, where $\varepsilon \ll 1$. It
is follows
\begin{eqnarray*}
 u^2(\xi) &=& \lim_{\varepsilon \to 0}\frac{4}{\vartheta}
    \left(\frac{2-\varepsilon/2}{1+2(2-\varepsilon/2)(\xi -
    \xi_0)^2}\right) =\\
    ~&=& \frac{8}{\vartheta[1+4(\xi - \xi_0)^2]}.
\end{eqnarray*}

Near right boundary of the interval for permissible value of
parameter $\delta$ by the use the substitution $\delta=
2-\varepsilon$. It results in following expression
\[u^2(\xi) = \lim_{\varepsilon \to 0}\frac{4}{\vartheta}
\left(\frac{\varepsilon/2}{1+\varepsilon(\xi -
    \xi_0)^2}\right) = 0. \]

Thus, when parameter $\delta$ runs to $-2$ the solutions of the
equations (\ref{k:m:28:1})-(\ref{k:m:28:3}) take the form of the
algebraic soliton
\begin{eqnarray}
  a_1^2(\xi) &=& \frac{8}{\vartheta(1+\beta_1)\{1+4(\xi -
\xi_0)^2\}}, \label{k:m:solito:alg:1}\\
  a_2^2(\xi) &=& \frac{8}{\vartheta(1-\beta_1)\{1+4(\xi -
\xi_0)^2\}}. \label{k:m:solito:alg:2}
\end{eqnarray}
Here $\beta_1$ corresponds with $\delta=-2$. If  $\delta$ runs to
$+2$, the amplitudes of the waves (\ref{k:m:28:1})-(\ref{k:m:28:3})
are equal to zero.

\section{Limitation of the velocity parameter $\beta$}

\noindent The velocity parameter $\beta$ is limited due to condition
 $|\delta|\leq 2 $. With the definition $$ \delta =
\left(\frac{\gamma_3}{\gamma_1}
\sqrt{\frac{1+\beta}{1-\beta}}+\frac{\gamma_2}{\gamma_1}
\sqrt{\frac{1-\beta}{1+\beta}}\right), $$ we can introduce function
\[F(\mu) = \frac{\gamma_3}{\gamma_1} \mu
+\frac{\gamma_2}{\gamma_1} \frac{1}{\mu},
\] depending on
\[\mu = \sqrt{\frac{1+\beta}{1-\beta}}. \]
It is odd function of $\mu$. Hence the condition $|\delta| \leq 2$
will be held in intervals $\mu_{-}\leq \mu\leq \mu_{+}$ and
$-\mu_{-}\geq \mu \geq -\mu_{+}$, where $\mu_{-}$ and $\mu_{+}$ are
positive roots for equation $|F(\mu)| = 2$. For positive $\mu$ if
$\min F(\mu)\leq 2$ the real roots exist. We can find that the
function $F(\mu)$ attains extremum at $\mu = \pm
\sqrt{\gamma_2/\gamma_3}$. If $\mu$ is positive value then this
extremum is minimum, else if $\mu$ is negative one the extremum is
maximum. The value of the function $F(\mu)$ at $\mu = +
\sqrt{\gamma_2/\gamma_3}$ is $2\sqrt{\gamma_2\gamma_3}$. Thus the
solitary waves are exist if $\sqrt{\gamma_2\gamma_3} \leq
|\gamma_1|$. From the definitions of parameters  $\gamma_{1,2,3}$ it
results in $\gamma_2\gamma_3 = 4c_2c_3 \cos^22\phi$. In discussed
model the coupling coefficients $c_{1,2,3}$ can be made positive
ones. Hence, the product $\gamma_2\gamma_3$ is positive too, or is
equal to zero at $\phi = \pi/4$.

Parameter $\delta$ can be represented as
$$ \delta = \left(\frac{c_2}{c_1}
\sqrt{\frac{1+\beta}{1-\beta}}+\frac{c_3}{c_1}
\sqrt{\frac{1-\beta}{1+\beta}}\right)\frac{\cos 2\phi}{\cos\phi}. $$
It follows that the phase shift for complex envelopes in neighboring
waveguides defines the $\sign\delta$. For example parameter $\delta$
will be negative one in intervals $\pi/2 <\phi < \pi$ and $3\pi/2
<\phi < 2\pi$.

If $\mu>0$ the roots of equation $F(\mu) = 2$ take the following
form
\begin{eqnarray*}
  \mu_{-} &=& \frac{|\gamma_1|}{|\gamma_2|}\left(1 -\sqrt{1-\frac{\gamma_2\gamma_3}{\gamma_1^2}}\right), \\
  \mu_{+} &=& \frac{|\gamma_1|}{|\gamma_2|}\left(1 +\sqrt{1-\frac{\gamma_2\gamma_3}{\gamma_1^2}}\right)
\end{eqnarray*}

The boundary of interval for permissible values of the velocity
parameters $\beta_{\pm}$ are determined by the expression
\[\beta_{\pm}=\frac{\mu_{\pm}^2 -1}{\mu_{\pm}^2 +1}.\]
For example, let us assume that $\gamma_2=\gamma_3 = 0,5 \gamma_1$.
In this case we have
\begin{eqnarray*}
  &\mu_{+} \approx 3,73  & \beta_{+} \approx 0,866,\\
  &\mu_{-} \approx 0,266 & \beta_{-} \approx -0,865
\end{eqnarray*}

It should be pointed that if the next nearest neighboring waveguides
are not coupled as in \cite{KMO:09}, then steady state waves exist
for $|\beta| < 1$.

\section {Numerical simulation of the solitary waves collisions}

\noindent To demonstrate stability of the solitary waves
corresponding with the system of equations (\ref{km:42:1}) and
(\ref{km:42:2}) the  collision between two solitary waves has been
simulated. The real envelopes were taken as (\ref{k:m:solito:1} and
\ref{k:m:solito:2}). The real phases corresponding them were chosen
as following ones
\begin{eqnarray}
\varphi_1(\zeta_j,\tau)&=& ~~\frac{3}{2}\mathrm{sgn}(\vartheta)
\mathcal{S}(0,2\xi_j ), \notag\\
\varphi_2(\zeta_j,\tau)&=& -\frac{\pi}{2} +
\frac{1}{2}\mathrm{sgn}(\vartheta) \mathcal{S}(0,2\xi_j),
\label{k:m:phase:nri:pri}
\end{eqnarray}
where $\xi_j=\gamma_1[\zeta_j-\beta(\tau-\tau_j)](1-\beta^2)^{1/2}$
The arguments $\zeta_j$ è $\tau_j$ define the initial position of
the $j$-th solitary wave.

To produce a collision between two solitary waves we used solutions
of the equations (\ref{km:42:1}) and (\ref{km:42:2}) with parameters
$\beta=0.4$ and $\beta=-0.4$ as initial conditions for these
equations. The pulse with  $\beta=+0.4$ was located at $\zeta=50$
and pulse with $\beta=-0.4$ was located at  $\zeta=0$. The coupling
constant $\gamma_1$ was set as unite. For simplicity the coupling
constants $\gamma_2$ and $\gamma_3$ are assumed equal, i.e.,
$\gamma_2=\gamma_3$.

\begin{figure}
\includegraphics[width=0.23\textwidth,scale=0.5]{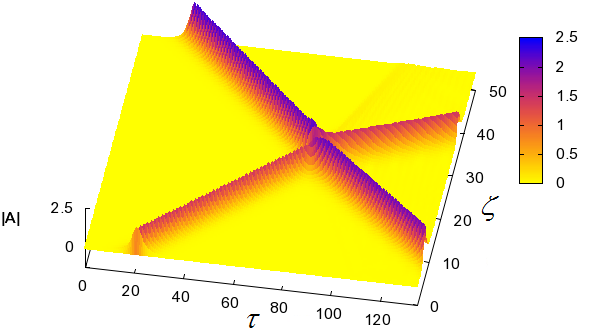} \
\includegraphics[width=0.23\textwidth,scale=0.5]{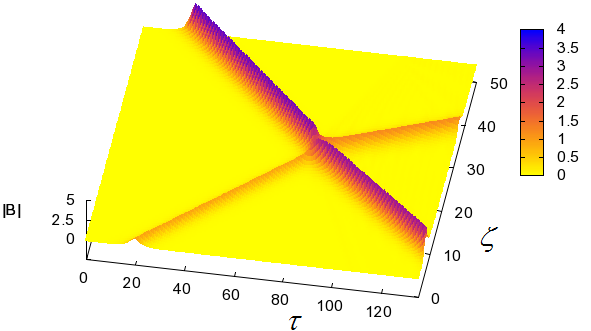}
\caption{{\protect\footnotesize {\ Collision of two steady state
pulses in the case of $\gamma_2=\gamma_3 =0$.}}}
\label{zigzagz:Fig_2}
\end{figure}

The case of the linear chain of the alternating refractive indexes
waveguide array corresponds to waveguide system with
($\gamma_2=\gamma_3=0$). Collision between two solitary waves is
represented in Fig.\ref{zigzagz:Fig_2} (see also \cite{KMO:09}).
Figures show that collision is elastic.

\begin{figure}
\includegraphics[width=0.23\textwidth,scale=0.5]{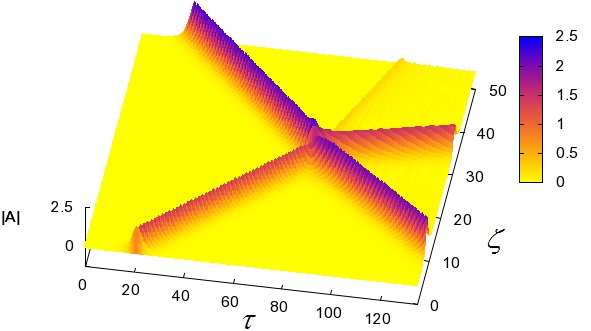} \
\includegraphics[width=0.23\textwidth,scale=0.5]{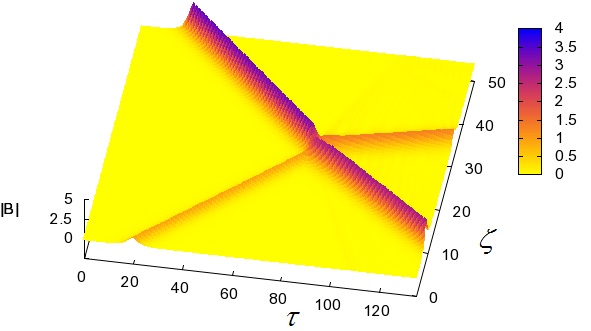}
\caption{{\protect\footnotesize {\ Collision of two steady state
pulses in the case of $\gamma_2=\gamma_3 =0.01$.}}}
\label{zigzagz:Fig_3}
\end{figure}

We found that the collision between counter propagating pulses is elastic
for the coupling constants $\gamma_2=\gamma_3$ that is taken from
interval $[0.001,~ 0.0075]$. Little radiation appears where the
coupling constants are taken from interval $[0.0075,~0.01]$
Fig.\ref{zigzagz:Fig_3}. For the coupling constants are from
interval $[0.001,~0.021]$ the amplitudes of the radiation are
increasing.

\begin{figure}
\includegraphics[width=0.23\textwidth,scale=0.5]{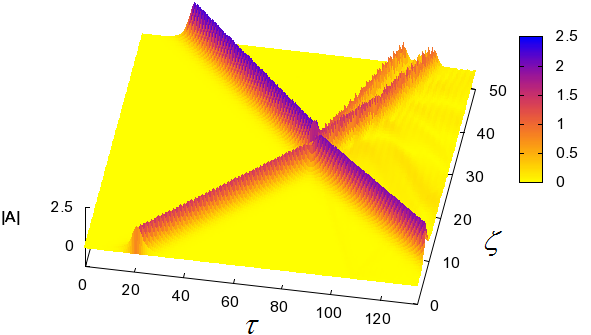} \
\includegraphics[width=0.23\textwidth,scale=0.5]{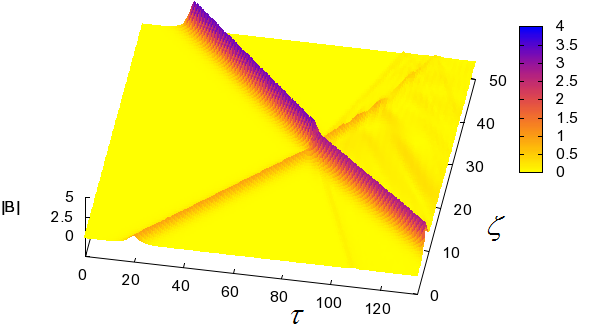}
\caption{{\protect\footnotesize {\ Collision of two steady state
pulses in the case of $\gamma_2=\gamma_3 =0.022$.}}}
\label{zigzagz:Fig_4}
\end{figure}

Increasing the coupling constants $\gamma_2=\gamma_3$, that are
belong to interval $[0.0215, ~ 0.03]$, leads to the transferred
wave decreasing as it is shown in Fig. \ref{zigzagz:Fig_4}.

\begin{figure}
\includegraphics[width=0.23\textwidth,scale=0.5]{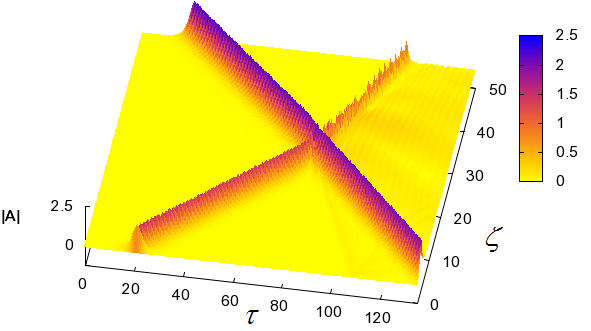} \
\includegraphics[width=0.23\textwidth,scale=0.5]{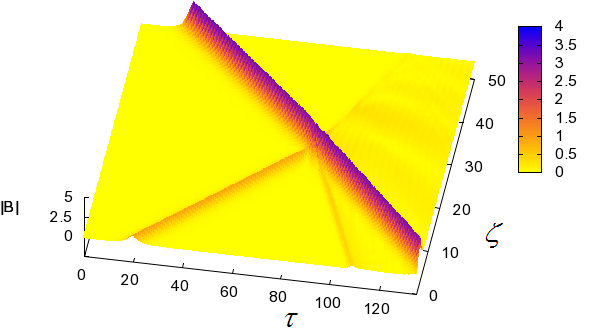}
\caption{{\protect\footnotesize {\ Collision of two steady state
pulses in the case of $\gamma_2=\gamma_3 =0.06$.}}}
\label{zigzagz:Fig_5}
\end{figure}

\begin{figure}
\includegraphics[width=0.23\textwidth,scale=0.5]{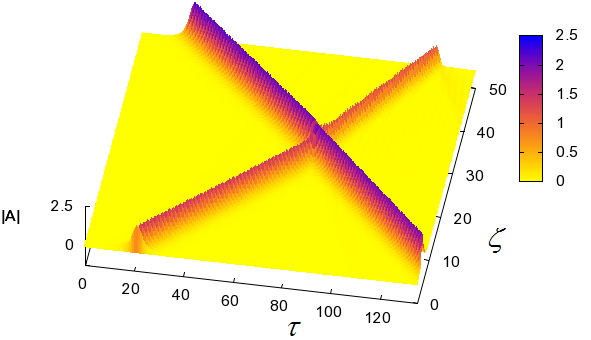} \
\includegraphics[width=0.23\textwidth,scale=0.5]{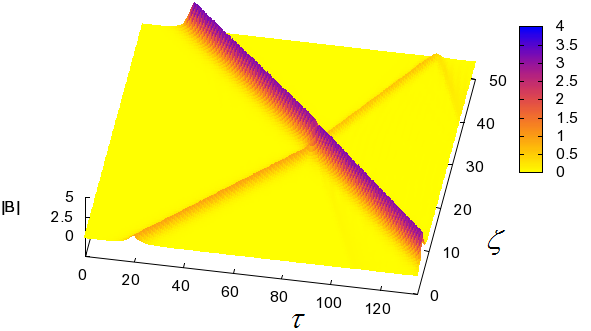}
\caption{{\protect\footnotesize {\ Collision of two steady state
pulses in the case of $\gamma_2=\gamma_3 =0.08$.}}}
\label{zigzagz:Fig_6}
\end{figure}

When à $\gamma_2=\gamma_3 \geq 0.02$ the reflected wave appears in
NRI waveguide as a result of the reflection of the incident solitary
wave ($\beta=-0.4$) from solitary wave ($\beta=0.4$) propagated in
the opposite direction in ANOWZA. The reflected wave amplitude grows
up to some value and thereafter it decrease in interval $[0.025, ~
0.05]$, Fig.\ref{zigzagz:Fig_5} and Fig.\ref{zigzagz:Fig_6}. The
reflected wave is disappeared at $\gamma_2=\gamma_3\approx 0.075$.

\begin{figure}
\includegraphics[width=0.23\textwidth,scale=0.5]{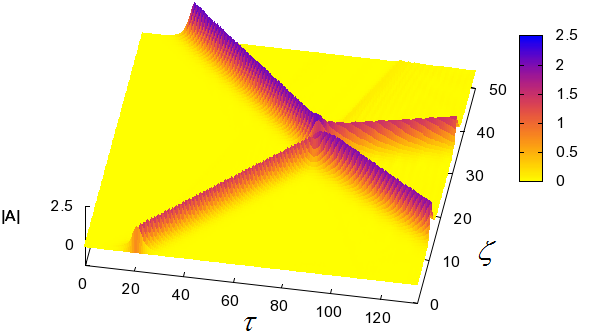} \
\includegraphics[width=0.23\textwidth,scale=0.5]{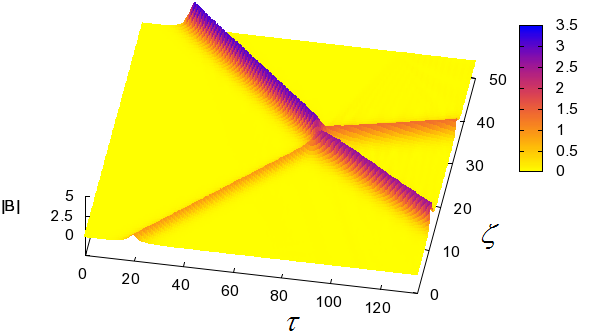}
\caption{{\protect\footnotesize {\ Collision of two steady state
pulses in the case of $\gamma_2=\gamma_3 =0.13$.}}}
\label{zigzagz:Fig_7}
\end{figure}

If the coupling constants belong to interval $[0.08, ~ 0.135]$ the
steady state solitary waves are akin to elastic interacting waves.
There is no radiation after collision, however the velocities of resulting waves
vary considerably. See Fig. \ref{zigzagz:Fig_7} and Fig.
\ref{zigzagz:Fig_2}.

\begin{figure}
\includegraphics[width=0.23\textwidth,scale=0.5]{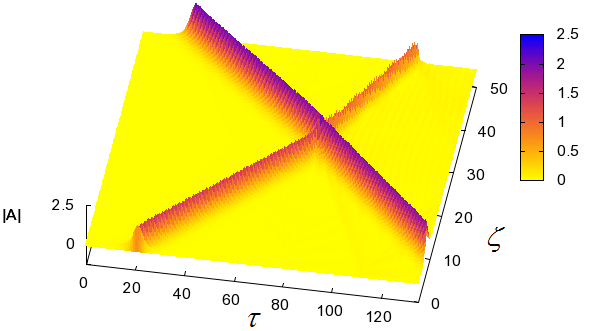} \
\includegraphics[width=0.23\textwidth,scale=0.5]{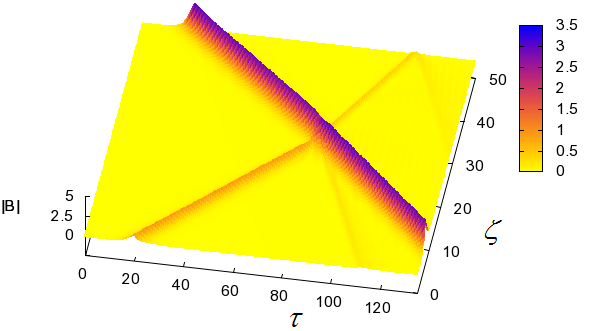}
\caption{{\protect\footnotesize {\ Collision of two steady state
pulses in the case of $\gamma_2=\gamma_3 =0.2$.}}}
\label{zigzagz:Fig_8}
\end{figure}

If $\gamma_2$ and $\gamma_3$ are taken from interval $[0.138, ~
0.15]$, the weak linear wave in NRI waveguide is generated. This
linear wave is absorbed by nonlinear transferred wave when
$\gamma_2$ and $\gamma_3$ rank among the interval $[0.151, ~ 0.16]$.
Reflected wave appear again when $\gamma_2$ and $\gamma_3$ are among
the intervals $[0.17, ~ 0.2]$ and $[0.31, ~ 0.34]$, see Fig.
\ref{zigzagz:Fig_8}

The steady state pulses collision does not generate the radiation if
the coupling constants $\gamma_2=\gamma_3$ are taken from interval
$[0.21, ~ 0.3]$. However the velocities of the scattered pulses can
be strongly varied with respect of initial values.

Also we had considered the interaction of the incident solitary wave
corresponding to $\beta=0.4$ with quasi-harmonic wave. Initial
incident pulse was located at $\zeta=50$, the quasi-harmonic wave
was generated at $\zeta=0$. This wave is described by following
expression
\begin{equation}
f(\tau)=f_0[\tanh(\tau-40)-\tanh(\tau-70)]\sin \omega_{bg} \tau.
\label{k:m:plateau}
\end{equation}
where $\omega_{bg}$ is the frequency of modulation, $f_0$ is the
amplitude of this plateau like pulse. Numerical simulations have
been proposed for $0.1 \leq f_0 \leq 3$ and $0.7\leq \omega_{bg}
\leq 3$.

\begin{figure}
\includegraphics[width=0.23\textwidth,scale=0.5]{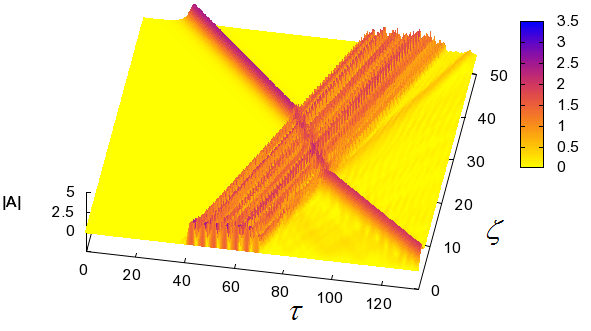} \
\includegraphics[width=0.23\textwidth,scale=0.5]{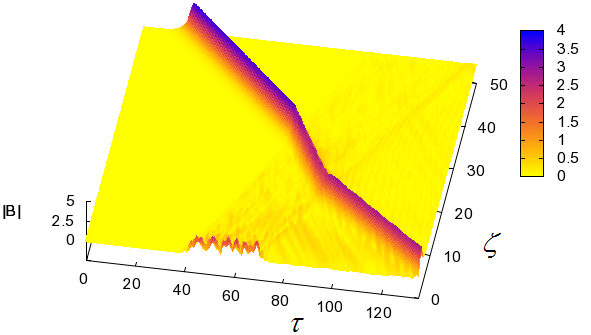}
\caption{{\protect\footnotesize {\ Steady state pulses crossing the
quasi-harmonic plateau like wave, $\gamma_2=\gamma_3 =0$.}}}
\label{zigzagz:Fig_9}
\end{figure}

\begin{figure}
\includegraphics[width=0.23\textwidth,scale=0.5]{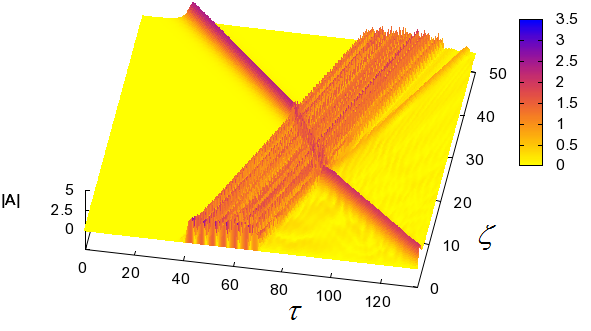} \
\includegraphics[width=0.23\textwidth,scale=0.5]{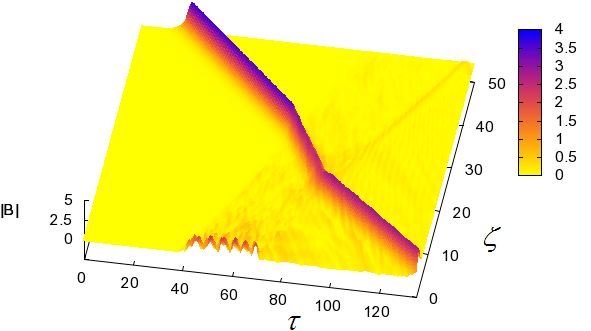}
\caption{{\protect\footnotesize {\ Steady state pulses crossing the
quasi-harmonic plateau like wave, $\omega_{bg}=0.7$,
$\gamma_2=\gamma_3 =0.01$.}}} \label{zigzagz:Fig_10}
\end{figure}

\begin{figure}
\includegraphics[width=0.23\textwidth,scale=0.5]{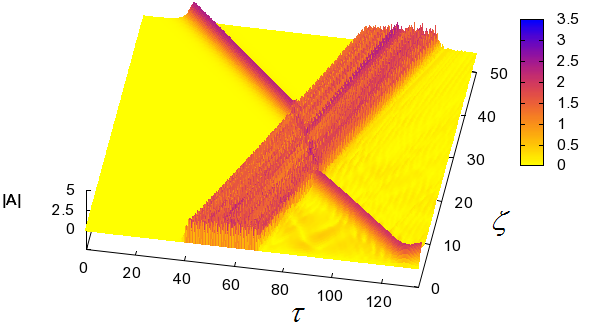} \
\includegraphics[width=0.23\textwidth,scale=0.5]{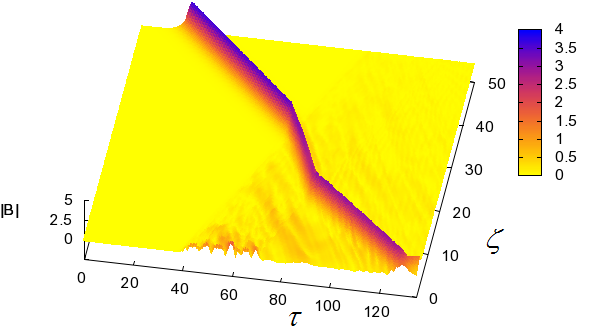}
\caption{{\protect\footnotesize {\ Steady state pulses crossing the
quasi-harmonic plateau like wave, $\omega_{bg}=3$,$\gamma_2=\gamma_3
=0.01$.}}} \label{zigzagz:Fig_11}
\end{figure}

It was found that initial steady state solitary wave is extremely
prone to damage if the coupling constants $\gamma_2=\gamma_3$ are
less than $0.05$. Fig.\ref{zigzagz:Fig_9} shows the propagation of
incident pulse trough quasi-harmonic wave (\ref{k:m:plateau}) at
$f_0=1.5$ and $\omega_{bg}=0.7$ and at $\gamma_2=\gamma_3=0$. For the
comparison the propagation of incident pulse trough the same
quasi-harmonic wave at $\gamma_2=\gamma_3=0,01$ is shown in
Fig.\ref{zigzagz:Fig_10}. There is no fundamental difference
between this case and the case of more high frequency quasi-harmonic wave
Fig. \ref{zigzagz:Fig_11}.

\section{Conclusion}

\noindent The propagation of electromagnetic solitary wave in zigzag
positive-negative coupled waveguide array is investigated. Coupling
between both nearest and next nearest neighboring waveguides is the
characteristic feature of this waveguide system. Alternate
positive and negative refractive indexes result in the gap in linear
waves spectrum. That is dissimilarity from a convenient waveguide
array.

The nonlinear waves were considered under assumption that NRI
waveguides are linear ones and PRI waveguides are manufactured from
the cubic nonlinear dielectric. Furthermore, we took hypothesis that
the phase difference between neighboring waveguides is constant. It
allows to reduce the infinite system of equations to system of two
equations and to find the steady state solutions of them.

The characteristic coupling constant for ANOWZA can be controlled by
the angle between the lines connecting the centers of neighboring
waveguides $\vartheta_b$. The numerical simulation of the collision
between two steady state solitary waves shows that complicated
picture of the interaction is depended on coupling constants which
are characterized coupling between the next nearest neighboring
waveguides. Thus the angle $\vartheta_b$ is control parameter for
nonlinear waves in ANOWZA.

{\bf Acknowledgements} \\
The research was partially supported by RFBR (grant No.
12-02-00561).


\begin{thebibliography}{999}

\bibitem{Bergamin:08} L. Bergamin,
Phys.Rev. \textbf{A78}, 043825 (12 pages) (2008).

\bibitem{Urzhumo:Smith:10} Ya.A. Urzhumov, D.R. Smith,
Phys.Rev.Lett. \textbf{105}, 163901 (2010).

\bibitem{KGSmith:10} N. Kundtz, D. Gaultney, D.R. Smith,
New J. Phys. \textbf{12}, 043039 (2010).

\bibitem{Shal:Pendry:11} Vl.M. Shalaev, J. Pendry,
J.Opt. \textbf{13}, 020201 (2011).

\bibitem{Kild:Shal:11} A.V. Kil'dishev, V.M. Shalaev,
Usp.Phys.Nauk, \textbf{181}, 59 (2011).

\bibitem{Diedrich:12} D. Diedrich, A. Rottler, D. Heitmann, S.
Mendach, New J. Phys. \textbf{14}, 053042 (2012)

\bibitem{Ch:Wu:Kong:06} H. Chen,
B.-I. Wu, and J. A. Kong
J.Electromagn. Waves and Appl. \textbf{20}, 2137 (2006).

\bibitem{Boltasseva:08} Al. Boltasseva, Vl.M. Shalaev,
Metamaterials \textbf{2}, 1 (2008).

\bibitem{Elefth:05}  \textit{Negative-refraction Metamaterials: Fundamental
Principles and Applications}, Eleftheriades G.V., Balmain K.G.
(Eds.) (New York: Wiley, 2005).

\bibitem{Noginov:12} Noginov M.A., Podolskiy V.A. (Eds) \textit{Tutorials in
Metamaterials} (Boca Raton – London – New York: Taylor and Francis
 Group, LLC/CRC Press, 2012).

\bibitem{Darmanyan:05} S.A. Darmanyan, M. Neviere, and A.A. Zakhidov,
Phys.Rev. E \textbf{72}, 036615 (2005)

\bibitem{Shadrirov:04} I.V. Shadrivov, Photonics Nanostruct. Fundam. Appl. \textbf{
2}, 175--180 (2004).

\bibitem{MGL:08}A.I. Maimistov, I.R. Gabitov , N.M. Litchinitser,
Optics and Spectroscopy \textbf{104}, 253 (2008).

\bibitem{KMO:09}  E.V. Kazantseva,  A.I. Maimistov, S.S. Ozhenko,
\textit{Phys.Rev. A} \textbf{80}, 43833 (7 pages) (2009).

\bibitem{LGM:07} N.M. Litchinitser, I.R. Gabitov, A.I. Maimistov,
Index Channel. Phys. Rev. Lett. \textbf{99}, ~113902 (2007).

\bibitem{Rizhov:Maim:12} M.S. Ryzhov, A.I. Maimistov,
Quantum Electronics. \textbf{42}, 1034--1038 (2012).

\bibitem{Christod:Leder:Silber:03}
D. N. Christodoulides, F. Lederer and Y. Silberberg,
Nature \textbf{424}, 817 (2003).

\bibitem{Christo:88} D.N. Christodoulides, R.I. Joseph, Opt.Letts. \textbf{13}, 794--796 (1988).

\bibitem{Sch:Lederer:91} C. Schmidt-Hattenberger, U. Trutschel, and F. Lederer,
Opt.Letts. \textbf{16}, 294--296 (1991).

\bibitem{Darmanyay:Leder:97} S. Darmanyan, I. Relke, and F. Lederer, Phys.Rev. E. \textbf{
55}, 7662--7668 (1997).

\bibitem{Efre:Chri:02} N.K. Efremidis, D. N. Christodoulides
\textit{Phys. Rev. B} \textbf{65}, 056607 (2002).

\bibitem{Wang:Huang:Yu:10} Gang Wang, Ji Ping Huang, and Kin Wah Yu
Optics Letters, \textbf{35}, 1908--1910 (2010).


\bibitem{Maim:Gabi:2007} A.~I. Maimistov, I.~R. Gabitov
\textit{Eur. Phys. J. Special Topics} \textbf{147}, 265 (2007).

\bibitem{Konotop:Abdullaev:12}
D.A. Zezyulin, V.V. Konotop, F.K. Abdullaev,
Optics Letters, \textbf{37}, 3930--3932 (2012).

\bibitem{Maim:Kazant:Desyat:12} A.I. Maimistov, E.V. Kazantseva,
A.S. Desyatnikov: Linear and nonlinear properties of the
antidirectional coupler, in Coherent optics and optical
spectroscopy, Lect.notes, Kazan State University, Kazan, (2102), pp.
21-31. (in Russian).



\end{thebibliography}
\end{document}